\documentclass[american,english,journal=jctcce,manuscript=article,layout=twocolumn]{achemso}
\usepackage[LGR,T1]{fontenc}
\usepackage[latin9]{inputenc}
\usepackage{array}
\usepackage{refstyle}
\usepackage{url}
\usepackage{bm}
\usepackage{multirow}
\usepackage{amsbsy}
\usepackage{graphicx}
\usepackage{esint}

\makeatletter

\title{Combining simulations and solution experiments as a paradigm for
RNA force field refinement }
\author{Andrea Cesari, Alejandro Gil-Ley, and Giovanni Bussi}
\affiliation{Scuola \foreignlanguage{american}{Internazionale} Superiore di Studi
Avanzati (SISSA), via Bonomea 265, 34136, Trieste, Italy}
\email{bussi@sissa.it}
\AtBeginDocument{\providecommand\eqref[1]{\ref{eq:#1}}}
\AtBeginDocument{\providecommand\figref[1]{\ref{fig:#1}}}
\AtBeginDocument{\providecommand\tabref[1]{\ref{tab:#1}}}
\DeclareRobustCommand{\greektext}{%
  \fontencoding{LGR}\selectfont\def\encodingdefault{LGR}}
\DeclareRobustCommand{\textgreek}[1]{\leavevmode{\greektext #1}}
\DeclareFontEncoding{LGR}{}{}
\DeclareTextSymbol{\~}{LGR}{126}
\providecommand{\tabularnewline}{\\}
\RS@ifundefined{subref}
  {\def\RSsubtxt{section~}\newref{sub}{name = \RSsubtxt}}
  {}
\RS@ifundefined{thmref}
  {\def\RSthmtxt{theorem~}\newref{thm}{name = \RSthmtxt}}
  {}
\RS@ifundefined{lemref}
  {\def\RSlemtxt{lemma~}\newref{lem}{name = \RSlemtxt}}
  {}

\usepackage{xr-hyper} 
\usepackage{hyperref}
\hypersetup{colorlinks=false,citecolor=black,hidelinks=true}

\AbstractOn
\let\oldmaketitle\maketitle
\let\maketitle\relax

\makeatother

\usepackage{babel}
\begin{document}
\maketitle
\twocolumn[ 
\begin{@twocolumnfalse} 
\oldmaketitle 
\begin{abstract} 
Recent computational efforts have shown that the current potential energy models used in molecular dynamics are not accurate enough to describe the conformational ensemble of RNA oligomers and suggest that molecular dynamics should be complemented with experimental data. We here propose a scheme based on the maximum entropy principle to combine simulations with bulk experiments. In the proposed scheme the noise arising from both the measurements and the forward models used to back calculate the experimental observables is explicitly taken into account. The method is tested on RNA nucleosides and is then used to construct chemically consistent corrections to the Amber RNA force field that allow a large set of experimental data on nucleosides and dinucleosides to be correctly reproduced. The transferability of these corrections is assessed against independent data on tetranucleotides and displays a previously unreported agreement with experiments. This procedure can be applied to enforce multiple experimental data on multiple systems in a self-consistent framework thus suggesting a new paradigm for force field refinement.
\end{abstract} 
\end{@twocolumnfalse} 
]

\section{Introduction}

Molecular dynamics (MD) simulations in explicit solvent have been
successfully applied to RNA systems in order to describe dynamics
around the native structure as well as small conformational changes.\cite{Banas2010,ColizziBussi2012,Yildirim2013,DiPalma2013,Musiani2014,Heldenbrand2014,SponeJPCL2014,GiambasuRNA2015,PinamontiNAR2015}
In the case of RNA folding, a few partly successful atomistic simulations
have been reported.\cite{ChenGarcia2013,KuehrovaBanasBestEtAl2013,Miner2016,Kuehrova2016}
However, recent extensive simulations of unstructured oligonucleotides
for which converged sampling is affordable have unambiguously shown
that current force-field parameters are not accurate enough to reproduce
solution experiments.\cite{CondonTurner2015,BergonzoCheatham2015,BottaroGil-LeyBussi2016,Gil-LeyBottaroBussi2016,bottaro2016free}
These results prompt for a new effort in RNA force field refinement
and suggest that MD should be complemented with experimental data,
when available. An important point here is the realization that many
experimental techniques intrinsically provide values that are averaged
both over time and over a large ensemble of replicates of the same
molecule. In this respect, many ways to enforce averages that match
experiments have been proposed in the past\cite{TordaScheekGunsteren1989,TordaVGunsteren1993,BestVendruscolo2004,LindorffLarsenBestVendruscolo2005,CavalliCamilloniVendruscolo2013,RouxWeare2013}.
The maximum entropy (MaxEnt) principle allows ensemble averages to
be constrained and is a natural framework to achieve this goal\cite{Jaynes1982,PiteraChodera2012}.
For instance, White and Voth implemented the MaxEnt procedure in their
experimentally directed simulation method (EDS)\cite{WhiteVoth2014}.
On top of this, recent works have underlined the importance of taking
into account experimental errors when using these constraints\cite{Boomsma2014,Hummer2015,BonomiMETAINF2016}
. This latter step is often done in a Bayesian framework, which is
not designed to provide a single optimal model but to sample the posterior
distribution thus generating ensembles of solutions compatible with
the constraints. 

In this work, we propose a novel method to enforce experimental data
in MD simulations. The method is an extension of the MaxEnt framework
and includes auxiliary variables to model experimental errors. Being
based on MaxEnt, this procedure allows a modified potential energy
function to be estimated on the fly. This method can be straightforwardly
combined with replica-exchange methods so as to enhance sampling and
improve the convergence of the modified potential. We apply the method
to the difficult task of designing a RNA force field that matches
NMR data from $^{3}J$ scalar couplings. The force field is restrained
to have a consistent functional form where chemically equivalent atoms
are subject to equivalent terms. The method is trained on a set of
nucleosides and dinucleoside monophosphates where it can reproduce
solution experiments with a previously unreported accuracy. Finally,
the obtained corrections are combined with a recently proposed correction
on the RNA backbone \cite{Gil-LeyBottaroBussi2016} and validated
using converged ensembles of tetranucleotides, where they significantly
improve the agreement with independent solution experiments, including
$^{3}J$ scalar couplings and nuclear-Overhauser-effect (NOE) data.
The success of the procedure suggests a new paradigm for force field
refinement.

\section{Method}

\subsection{MaxEnt approach}

The rationale behind the maximum entropy  (MaxEnt) approach, as firstly
proposed by Jaynes\cite{Jaynes2}, is to find a normalized probability
distribution $P(\boldsymbol{x})$ which satisfies a set of constraints
and minimizes the Kullback\textendash{}Leibler divergence\cite{Kullback1951}
from a reference probability $P_{0}(\boldsymbol{x})$, defined as:
\begin{equation}
D_{KL}(P||P_{0})=\int d\boldsymbol{x}P(\boldsymbol{x})\ln\frac{P(\boldsymbol{x})}{P_{0}(\boldsymbol{x})},\label{eq:kldivergence}
\end{equation}
Here $P_{0}(\boldsymbol{x})$ expresses our prior information on the
system and should be interpreted as the canonical distribution associated
with the initial, unrefined potential energy function. Minimizing
$D_{KL}$ is equivalent to maximizing the relative entropy between
$P$ and $P_{0}$ and, by construction, provides an ensemble that
introduces the minimum possible amount of information with respect
to the prior knowledge. The ensemble averages of a set of $M$ observables
$f_{i}(\boldsymbol{x})$ $i=1,\dots,M$, in the refined ensemble $P(\boldsymbol{x})$,
are equal to $\langle f_{i}\rangle=\int d\boldsymbol{x}P(\boldsymbol{x})f_{i}(\boldsymbol{x})$
and should be constrained to their experimental values $f_{exp,i}$:

\begin{equation}
f_{exp,i}=\int d\boldsymbol{x}P(\boldsymbol{x})f_{i}(\boldsymbol{x}).\label{eq:obs-constraint}
\end{equation}
Here $f_{i}(\boldsymbol{x})$ is a function of the atomic coordinates
of the system and depends on the nature of the performed experiment.
For example, three-bond scalar couplings $^{3}J(\theta)$ are related
to the dihedral angles formed by the $4$ involved atoms according
to the Karplus relations\cite{Karplus1963}: $f(\theta)={}^{3}J(\theta)=Acos^{2}\theta+Bcos\theta+C$.
Minimizing the functional in \eqref{kldivergence} using the method
of Lagrangian multipliers leads to: 

\begin{equation}
P(\boldsymbol{x})=\frac{P_{0}(\boldsymbol{x})e^{-\sum_{i=1}^{M}\lambda_{i}f_{i}(\boldsymbol{x})}}{\int d\boldsymbol{x}P_{0}(\boldsymbol{x})e^{-\sum_{i=1}^{M}\lambda_{i}f_{i}(\boldsymbol{x})}}.
\end{equation}
In the rest of the paper we will refer to this distribution as posterior
distribution. Here $\lambda$ is an array of Lagrangian multipliers
that should be determined in a self-consistent procedure. This corresponds
to a refined potential equal to:

\begin{equation}
U(\boldsymbol{x})=U_{0}(\boldsymbol{x})+k_{B}T\sum_{i=1}^{M}\lambda_{i}f_{i}(\boldsymbol{x})\label{eq:maxent_potential}
\end{equation}

It has been shown by Voth et al.\cite{DannenhofferLafage2016} that
such functional form of the biasing potential will always reduce the
relative entropy with respect to an ideal ensemble, in which all observable
are in agreement with experiments.

To determine the Lagrangian multipliers it is convenient\cite{Mead1984,PiteraChodera2012}
to define a function $\Gamma(\lambda)=\ln\int d\boldsymbol{x}e^{-\beta U(\boldsymbol{x})}+\sum_{i=1}^{M}\lambda_{i}f_{exp,i}$.
The gradient of $\Gamma$ is

\begin{equation}
\frac{\partial\Gamma}{\partial\lambda_{i}}=f_{exp,i}-\langle f_{i}(\boldsymbol{x})\rangle.
\end{equation}
Since for uncorrelated observables the Hessian of $\Gamma$ is positive
definite, a set of Lagrangian multipliers $\{\lambda^{*}\}$ satisfying
the constraints in \eqref{obs-constraint} can be found by minimizing
$\Gamma$. In the case of correlated observables the function \textgreek{G}
becomes positive semidefinite changing its character from strictly
convex to convex. This implies that there could exist multiple choices
of $\left\{ \lambda^{*}\right\} $ satisfying the constraints. It
can be shown however that if the constraints are satisfied by different
set of $\left\{ \lambda^{*}\right\} $, they all can be considered
as acceptable solutions to the problem (see Supporting Information
subsection 1.1). 

The optimal values for $\lambda$ can be computed on the fly during
a MD simulation using a stochastic procedure. Similarly to Ref. \cite{WhiteVoth2014},
we use a stochastic gradient descent method where $\lambda$ is updated
according to

\begin{equation}
\dot{\lambda}_{i}(t)=-\eta_{i}(t)\left(f_{exp,i}-f_{i}(\boldsymbol{x}(t))\right)\label{eq:update-rule-no-error}
\end{equation}
Here $\eta$ is a suitable learning rate which we chose from the class
type ``search then converge'' as $\eta(t)=\frac{k_{i}}{1+t/\tau_{i}}$
following Ref\cite{darken1991}. Here $k_{i}$ represents the initial
learning rate and $\tau_{i}$ represents its damping time. In this
manner, the learning rate is large at the beginning of the simulation
and decreases proportionally to $1/t$ for large simulation times.
Interestingly, a $1/t$ schedule for the learning rate has been found
to be optimal also in the context of enhanced sampling methods \cite{BarducciBussiParrinello2008}.
The simulation can be stopped when Lagrangian multipliers are converged.
If required, the simulation can be continued using a static correcting
potential. The best estimate of the values of $\lambda$ satisfying
the experimental constraints is given by the time average $\lambda_{i}^{*}$
of the Lagrangian multipliers over an appropriate time window $[t_{min},t_{max}]$.
In the rest of the paper, we call ``learning phase'' the initial
part of the simulation  ($t<t_{max})$, ``averaging phase'' the
portion of the learning phase where $\lambda$ is averaged  ($t_{min}<t<t_{max})$,
and ``production phase'' the later part of the simulation  ($t>t_{max}$),
where $\lambda$ is kept equal to the computed average. It should
be mentioned that the procedure used here is related to a recently
introduced a variationally enhanced sampling approach\cite{Valsson2014}
where a similar stochastic minimization is used to enforce target
distributions during a MD simulation.

\subsection{Inclusion of experimental errors}

The procedure outlined above is appropriate when available experimental
data can be assumed to be exact. However, in practical cases many
sources of error can affect the data, including systematic and random
errors on the data as well as errors arising from a sub-optimal parametrization
for the functions $f_{i}(\boldsymbol{x})$. In our example, the latter
case would correspond to a sub-optimal estimation of the parameters
in the Karplus relations. Requiring a perfect agreement with experiments
could lead to huge Lagrangian multipliers. We here extend the MaxEnt
formalism so as to explicitly account for all the above mentioned
errors. Namely, we consider an extended system where an additional
variables $\epsilon_{i}$ is introduced to take into account the deviation
between the experimental data $f_{exp,i}$ and the ensemble average
$\langle f_{i}\rangle.$ The enforced constraints are thus

\begin{equation}
\langle\left(f_{i}(\boldsymbol{x})+\epsilon_{i}\right)\rangle=f_{exp,i}.\label{eq:constraint_error}
\end{equation}
A priori, the variables $\epsilon_{i}$ are considered to be independent
of the atomic coordinates $\boldsymbol{x}$, and with a prior probability
distribution $P_{0}(\epsilon)$. To keep the notation simple, we denote
with the same letter both the prior on the atomic coordinates $P_{0}(\boldsymbol{x})$
and the prior on the deviation $P_{0}(\epsilon)$, as these two functions
can be easily distinguished by their argument. The prior $P_{0}(\epsilon)$
can be used to model the experimental errors. In the simplest case,
one can assume it to be a Gaussian function with a given standard
deviation $\sigma_{0}$, $P_{0}(\epsilon)=\frac{e^{-\frac{\epsilon^{2}}{2\sigma_{0}^{2}}}}{\sqrt{2\pi}\sigma_{0}}$.
Here a single experimental value has been assumed   ($M=1$), but
equations are straightforwardly generalized to multiple data. Different
values of $\sigma_{0}$ can be chosen for different types of experiment.
In more general cases, if $\sigma_{0}$ is difficult to choose, one
can consider it as  an additional variable with an appropriate prior
distribution $P_{0}(\sigma_{0})$. The resulting prior for $P_{0}(\epsilon)$
can thus be obtained by marginalization. 

Once the prior $P_{0}(\epsilon)$ has been defined, one should enforce
the constraint in \eqref{constraint_error}. This is done by applying
the MaxEnt procedure on the extended system defined by the coordinates
$\bm{x}$ and the auxiliary variables $\epsilon_{i}$. The resulting
posterior distribution is
\[
P(\boldsymbol{x,\epsilon})=\frac{P_{0}(\boldsymbol{x})P_{0}(\epsilon)e^{-\sum_{i}\lambda_{i}(f_{i}(\boldsymbol{x})+\epsilon_{i})}}{\int d\boldsymbol{x}d\epsilon P_{0}(\boldsymbol{x})P_{0}(\epsilon)e^{-\sum_{i}\lambda_{i}(f_{i}(\boldsymbol{x})+\epsilon_{i})}}
\]
As a consequence of this exponential form, $\bm{x}$ and $\epsilon$
are independent also in the posterior distribution and hence can be
factorized. The enforced constraint can thus be written as
\[
\left\langle f_{i}(\boldsymbol{x})\right\rangle +\left\langle \epsilon_{i}\right\rangle =f_{exp,i}
\]
The average value of $\epsilon$ depends only on the value of $\lambda$
and can be analytically computed without explicitly sampling $\epsilon$
as:
\[
\xi_{i}(\lambda)\equiv\left\langle \epsilon_{i}\right\rangle =\frac{\int d\epsilon P_{0}(\epsilon)e^{-\sum_{j}\lambda_{j}\epsilon_{j}}\epsilon_{i}}{\int d\epsilon P_{0}(\epsilon)e^{-\sum_{j}\lambda_{j}\epsilon_{j}}}
\]
where $\xi_{i}(\lambda)$ is defined as the average value of $\epsilon$
in the posterior distribution. Once the functional form of $\xi_{i}(\lambda)$
is known, one should enforce the average values obtained from the
MD simulation to be equal to $f_{exp,i}-\xi_{i}(\lambda)$. By applying
the stochastic minimization procedure described above to the extended
system one obtains the following update rule for $\lambda$
\begin{equation}
\dot{\lambda}_{i}(t)=-\eta(t)\left(f_{exp,i}-\xi_{i}(\lambda)-f_{i}(\boldsymbol{x}(t))\right)\label{eq:update-rule}
\end{equation}
This equation is the most important of this paper and completely describes
the algorithm that we use to restrain MD with noisy experimental data.
The treatment of the error is fully enclosed in the functional form
of $\xi_{i}(\lambda)$. The equations of motion are integrated using
the algorithm reported in Supporting Information (subsection 1.2),
where it is also explained how to include experiments providing only
an upper or lower limit for a given observable. Since $\lambda$ changes
during the simulation, the system is kept out of equilibrium. The
work performed can be estimated (see Supporting Information equation
1.2) and used to compute the effective energy drift\cite{FerrarottiBottaroPerez-VillaEtAl2015}.

In the simple case where $P_{0}(\epsilon)$ is a Gaussian with standard
deviation $\sigma$ the $\xi$ function is
\begin{equation}
\xi_{i}(\lambda)=-\lambda_{i}\sigma^{2}.\label{eq:xi-gaussian}
\end{equation}
Larger values of $\sigma$ would lead to smaller Lagrangian multipliers
at the end of the stochastic minimization. The value of $\sigma$
thus tells us how much we trust in the original force field and is
related to the value of $\theta$ introduced in Ref\cite{Hummer2015}.
Non Gaussian prior distributions for $\epsilon$ can be used to better
tolerate outliers. The Laplace prior $P_{0}(\epsilon)\propto e^{-\sqrt{2}\frac{\left|\epsilon\right|}{\sigma}}$
results in

\begin{equation}
\xi_{i}(\lambda)=-\frac{\lambda_{i}\sigma^{2}}{1-\frac{\lambda^{2}\sigma^{2}}{2}}\label{eq:xi-laplace}
\end{equation}
This function is similar to the one in \eqref{xi-gaussian} for small
values of $\lambda$. However, when $\lambda$ approaches the limiting
value of $\frac{\sqrt{2}}{\sigma}$, $\xi$ diverges and arbitrarily
large discrepancies with the experimental data are accepted. This
procedure forces $\lambda$ to be in the domain $\left[-\frac{\sqrt{2}}{\sigma},+\frac{\sqrt{2}}{\sigma}\right]$,
and thus intrinsically limits the strength of the corrections to the
original force field. We notice that these boundaries ensure that
the posterior $P(\epsilon)\propto e^{-\sqrt{2}\frac{\left|\epsilon\right|}{\sigma}}e^{-\lambda\epsilon}$
can be normalized.

A general class of priors $P_{0}(\sigma_{0})$ is presented in Supporting
Information (subsection 1.3) that comprises the above mentioned Gaussian
and Laplace priors $P_{0}(\epsilon)$ as special cases. We notice
that, although in this examples $\xi_{i}$ only depends on $\lambda_{i}$,
in principle any element of $\xi$ could depend on the value of any
of the Lagrangian multipliers. This could happen when errors on different
experimental data are considered as correlated (see Supporting Information
subsection 1.4).

\subsection{Self-consistent force-field refinement}

Once the set of $\{\lambda^{*}\}$ satisfying \eqref{constraint_error}
are determined, the potential energy used in the MaxEnt framework
(\eqref{maxent_potential}) is equivalent to the original force field
plus a correction linear in the experimental observables. Although
these corrections have been fitted to match experiments on one specific
system, one could try to transfer them to a system different from
the original one. This is particularly appealing in the case of $^{3}J$
couplings, where the functional form of the correction is comparable
to the standard torsional terms that are present in biomolecular force
fields. We propose here to transfer the parameters directly during
the learning phase. If experimental data are available for a number
of similar systems, one should simulate all the systems in parallel.
Each of the simulated systems will be affected both by the corrections
arising from the experiments performed on the same system and by some
of the correcting potentials determined by the other simulated systems.
This procedure allows to fit force-field corrections in a self-consistent
procedure that restrains some of the terms to be equivalent to each
other. In the case of corrections derived by $^{3}J$ couplings, it
is possible to enforce the same correction on dihedrals that are chemically
equivalent to each other. For instance, the torsional potential of
a $\chi$ angle in an adenine is expected to be the same irrespectively
of its position in the sequence. A detailed description of the procedure
can be found in Supporting Information  (subsection 1.5). The systems
considered are: A, C, ApA, ApC, CpA, and CpC. Experimental data for
the dinucleoside monophosphates were taken from Refs.\cite{LeeEzraKondoEtAl1976,EzraLeeKondoEtAl1977,SponerEtAl2009}

\subsection{Molecular dynamics }

We performed molecular dynamics on RNA nucleosides  (A and C) and
dinucleoside monophosphates  (ApA, ApC, CpA, and CpC). Molecular dynamics
simulations were performed using the GROMACS software package \cite{GROMACS2013}
in combination with a modified version of the PLUMED plugin\cite{PLUMED2_2014}.
RNA, explicit water, and ions were modeled using the most recent parametrizations
within the Amber force field\cite{Jorgensen1983,Cornell1995,Perez2007,Cheatham2008,Banas2010,ZgarbovaOtyepkaSvponerEtAl2011}.
Parameters are available at \url{http://github.com/srnas/ff}. Bonds
were constrained using the LINCS algorithm\cite{LINCS1997}, allowing
for a timestep of 2 fs. The particle-mesh Ewald algorithm\cite{DardenYorkPedersen1993}
was used for long-range electrostatic interactions with a cut-off
distance of $1$ nm. Simulations were performed at temperature T =
300 K and pressure P = 1 bar \cite{ParrinelloRahman1980,Bussi_v-rescale2007}.
To allow for a fast convergence of the simulated ensembles, sampling
was enhanced using replica-exchange with collective-variable tempering
 (RECT) \cite{Gil-LeyBussi2015} on selected collective variables.
For the nucleosides we biased the torsional angles $\chi$, $\gamma$,
and the puckering variables $Z_{x}$ and $Z_{y}$ \cite{HuangYork2014}.
For the dinucleoside monophosphates we additionally included torsional
angles $\alpha$, $\beta$, $\epsilon$, and $\zeta$ as well as the
distance between the two nucleobases. Four replicas were used for
each system, with bias factors ranging from 1 to $5$ both for the
nucleosides and dinucleoside monophosphates.

\subsection{MaxEnt algorithm parameters}

For the nucleosides we performed $200\ ns$ MD per replica using the
first $100\ ns$ as learning phase. Lagrangian multipliers were averaged
from $t_{min}=50\ ns$ to $t_{max}=100\ ns$ and these averages were
used in the production phase for the last $100\ ns$. For the dinucleoside
monophosphates we performed $600\ ns$ using first $300\ ns$ as learning
phase and averaging Lagrangian multipliers between $t_{min}=150\ ns$
and $t_{max}=300\ ns$.

The parameters for the learning phase were chosen as $k=0.001\ Hz^{-2}ps^{-1},\ \tau=3\ ps,\ \sigma=2.0\ Hz$
for both the nucleosides and the dinucleoside monophosphates. In both
cases a Laplace prior for the error was used. The biased replicas
were simulated using Lagrangian multipliers estimated on the fly from
the reference replica, so as to maximize the acceptance rate for the
replica-exchange procedure. To implement the self-consistent force-field
fitting described above, we simultaneously simulated six systems  (A,
C, ApA, ApC, CpA, and CpC). The replica exchange framework of GROMACS
was used, disallowing unphysical exchanges between replicas simulating
different systems. Each system was simulating with 4 RECT replicas,
resulting in a total of 24 replicas. Lagrangian multipliers were adjusted
to fit experimental data available for each of the systems and transmitted
on the fly to the other replicas so as to be applied on all the equivalent
dihedrals. Input files are provided in Supporting Information (figure
S5-S9). The modifications to PLUMED required to perform this simulations
are available on request and will be included in the next PLUMED release.

\section{Results}

In the following we show a number of applications of the discussed
method. First we enforce data from solution experiments on RNA nucleosides,
showing that the procedure can be used to construct an ensemble compatible
with experiments. Then, the self-consistent procedure is used to fit
force-field corrections for nucleosides and dinucleosides composed
of adenine and cytosine. The obtained corrections are validated on
the difficult case of RNA tetranucleotides. A model one-dimensional
system which is designed to highlight the features of the different
models for including the experimental errors is shown in Supporting
Information (see subsection 2.1)

\subsection{Enforcing $^{3}J$ coupling on a nucleoside}

\begin{figure}
\begin{centering}
\includegraphics[scale=0.5]{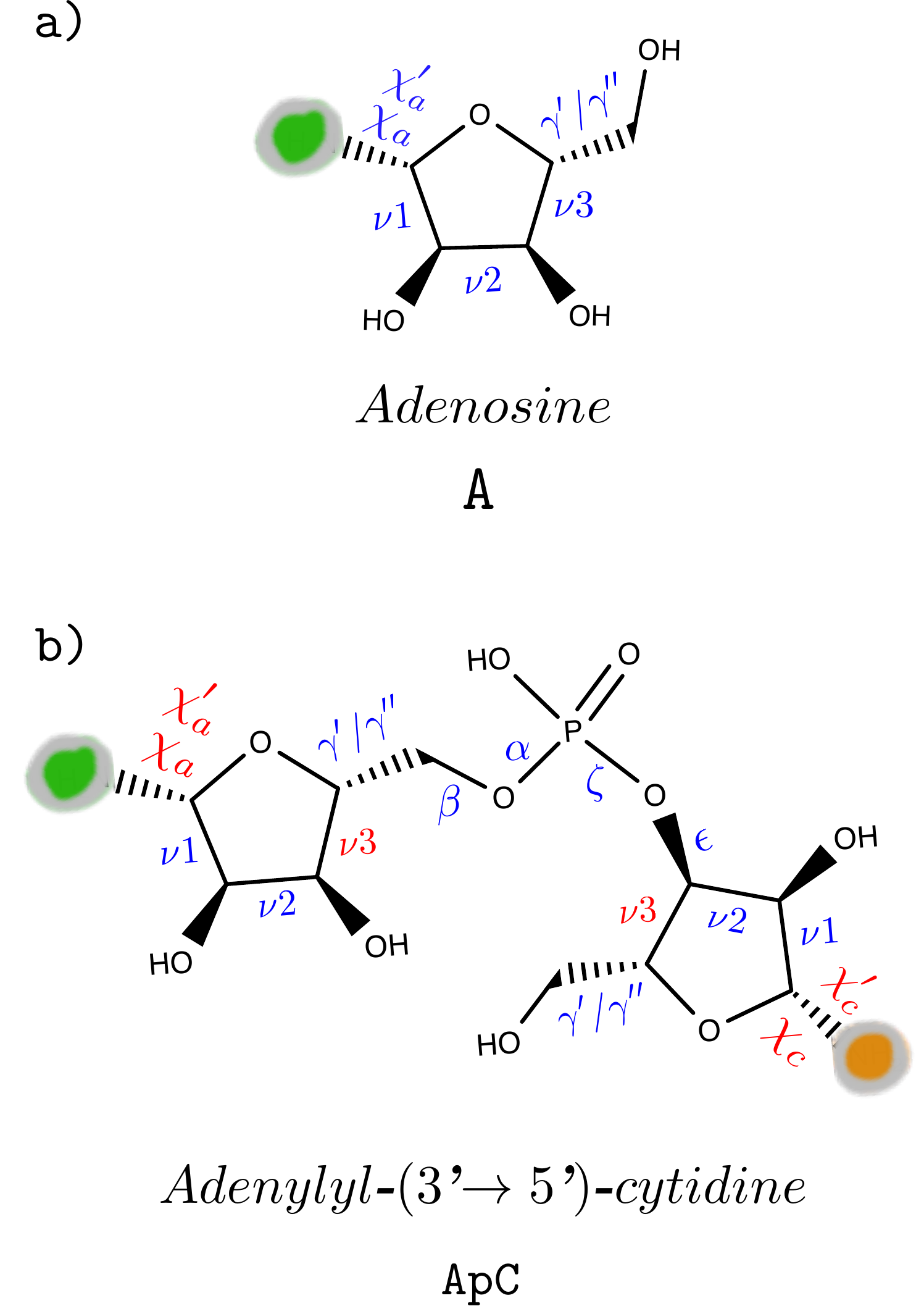}
\par\end{centering}

\caption{Torsional angles associated to the available experimental $^{3}J$
scalar couplings for the Adenosine (panel a) and the ApC dinucleoside
(panel b). Atoms associated to each torsion are: $\nu_{1}=$ H1'-C1'-C2'-H2',
$\nu_{2}=$ H2'-C2'-C3'-H3', $\nu_{3}=$ H3'-C3'-C4'-H4', $\gamma$'
= H4'-C4'-C5'-H5', $\gamma$'' = H4'-C4'-C5'-H5'', $\epsilon_{1}=$
C4'-C3'-03'-P, $\zeta_{1}=$ C3'-03'-P-05', $\alpha_{2}=$ 03'-P-05'-C5',
$\beta_{2}=$ P-05'-C5'-C4', $\chi_{A}=$ 04'-C1'-N9-C4, $\chi_{A}'=$
H1'-C1'-N9-C8+60\protect\textdegree, $\chi_{C}=$ O4'-C1'N1-C2,
$\chi_{C}'=$ H1'-C1'-N1-C6+60\protect\textdegree \label{fig:torsionals}}
\end{figure}

We show here an application of the introduced procedure to an RNA
nucleoside. We here discuss results for adenosine only. Results for
other nucleosides  (uridine, cytidine, and guanosine) are similar
and are summarized in Supporting Information  (table S1). For this
system, $M=7$ experimental $^{3}J$ scalar couplings are available\cite{Ancian2010},
involving dihedral angles both on the backbone and on the nucleobase
(see \figref{torsionals}a). 

We assess the deviation between simulation and experiments by computing
the RMSE of the back calculated data from their experimental values,
defined as $\sqrt{\frac{1}{M}\sum_{i=1}^{M}\left(^{3}J_{i,simulated}-^{3}J_{i,exp}\right)^{2}}$.
We first computed the scalar couplings using the standard Amber force
field  (see \tabref{-adenosine-scalar-coupling}). The RMSE in this
case is 1.3 Hz. This number is significantly larger than the expected
experimental error. However, it is important to consider also errors in the parametrization
of the Karplus equations. To this aim,
we compared a set of commonly used parametrizations (see Supporting Information subsection 1.6) and
computed their standard deviation on the trajectory corresponding to the ApC dinucleoside monophosphate
(reported in the next section),
which resulted in approximately 0.6 Hz. This number is significantly smaller than the RMSE observed for the
Amber force field.
This test also indicates that enforcing an RMSE between simulation and experiment
lower than 0.6 could lead to results dependent on the choice of the Karplus equation parameters.

Additionally, we estimated the ability of random conformations to reproduce the experimental
$^3$J scalar couplings. To this aim,
we computed the RMSE between simulation and experiments assuming a flat distribution on all the torsions used in the $^3$J coupling calculation. The torsions considered were the ones available for the ApC dinucleosides with the same set of Karplus parameters which was used to produce all the results in this work. The resulting RMSE is approximately 2.9 Hz, indicating that random conformations do not reproduce experimental data
with the accuracy of MD ensembles.

\begin{figure*}
\begin{centering}
\includegraphics[scale=0.6]{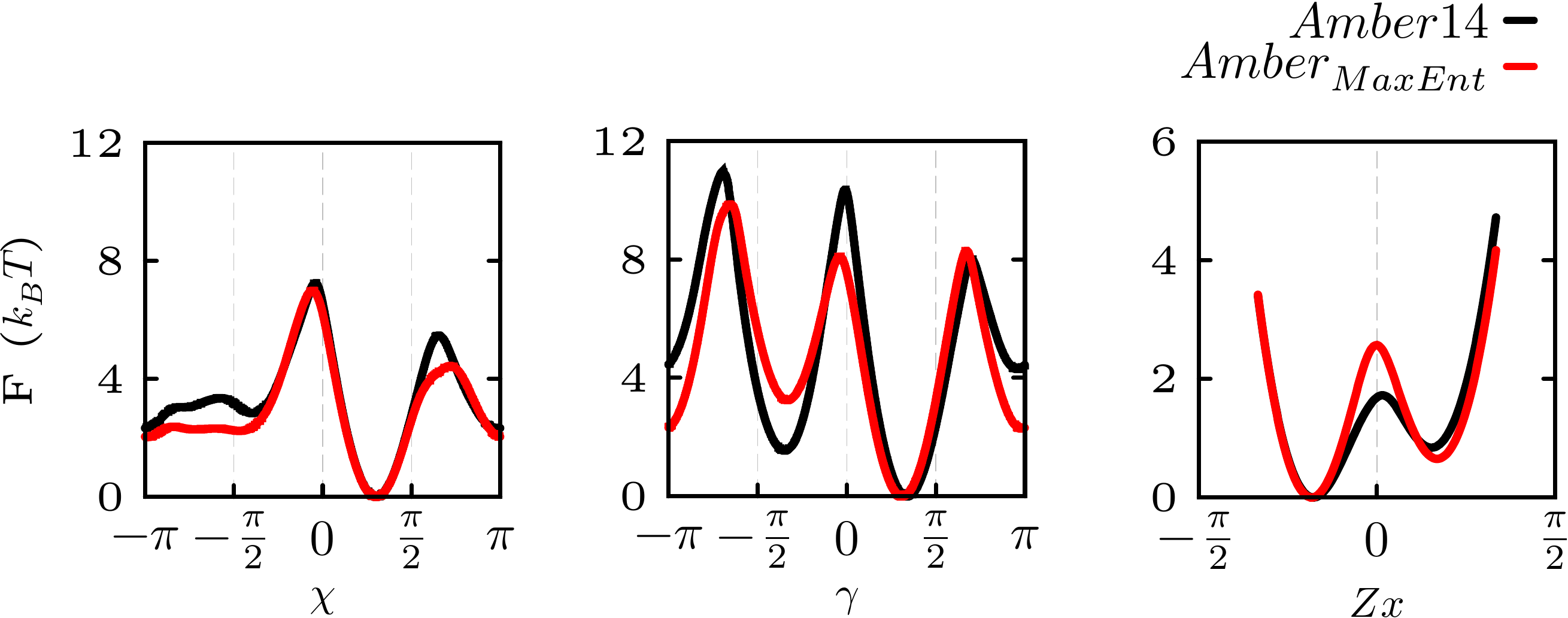}
\par\end{centering}

\caption{One-dimensional free-energy profiles for a representative group of
the corrected dihedral angles obtained with Amber and with the refined
Amber$_{MaxEnt}$ force fields. $Z_{x}$ variable\cite{HuangYork2014}
is related to sugar conformations C3'-endo ($Z_{x}>0)$ and C2'-endo
($Z_{x}<0$).\label{fig:adenosine-free-energy}}
\end{figure*}

We then
use our iterative procedure to determine the correcting potentials.
Although we use a Laplace prior for the error, we notice that since
the correcting potential has as many degrees of freedom as experimental
data, one cannot expect to detect inconsistencies in the dataset.
A crucial parameter in the fitting procedure is $\sigma$, which controls
the width of the prior distribution for the deviation between experiment
and theory, and encodes the confidence that we have in the force field.
Results for $\sigma=2.0$ Hz are shown in \tabref{-adenosine-scalar-coupling}.
\begin{table}
\begin{centering}
\begin{tabular}{|>{\centering}m{10mm}|>{\centering}m{1.8cm}|>{\centering}m{1.8cm}|>{\centering}m{1.8cm}|}
\cline{2-4} 
\multicolumn{1}{>{\centering}m{10mm}|}{} & \multicolumn{3}{c|}{{\small{}$^{3}J$ coupling $\left(Hz\right)$}}\tabularnewline
\hline 
{\scriptsize{}torsion} & {\scriptsize{}Exp.\cite{Ancian2010}} & {\scriptsize{}Amber } & {\scriptsize{}Amber$_{MaxEnt}$}\tabularnewline
\hline 
\hline 
$\nu_{1}$ & {\small{}6.0} & {\small{}8.5} & {\small{}6.9}\tabularnewline
\hline 
$\nu_{2}$ & {\small{}5.0} & {\small{}5.1} & {\small{}5.1}\tabularnewline
\hline 
$\nu_{3}$ & {\small{}3.4} & {\small{}3.5} & {\small{}4.2}\tabularnewline
\hline 
$\gamma$' & {\small{}3.0} & {\small{}3.2} & {\small{}3.1}\tabularnewline
\hline 
$\gamma$'' & {\small{}3.4} & {\small{}1.5} & {\small{}2.6}\tabularnewline
\hline 
{\scriptsize{}$\chi$} & {\small{}3.6} & {\small{}4.7} & {\small{}4.1}\tabularnewline
\hline 
{\scriptsize{}$\chi^{'}$} & {\small{}3.9} & {\small{}3.6} & {\small{}3.5}\tabularnewline
\hline 
\multicolumn{1}{>{\centering}m{10mm}|}{} & \multicolumn{3}{c|}{{\footnotesize{}$RMSE\;(Hz)$}}\tabularnewline
\cline{2-4} 
\multicolumn{1}{>{\centering}m{10mm}|}{} & {\small{}0.0} & {\small{}1.3} & {\small{}0.6}\tabularnewline
\cline{2-4} 
\end{tabular}
\par\end{centering}

\caption{$^{3}J$ scalar coupling for the Adenosine nucleoside. Experimental
values and back calculated values are shown, both using the Amber
force field and the MaxEnt corrections. Angle $\chi'$ for the Adenosine
nucleoside is defined as the $H1^{'}-C1^{'}-N9-C8$ torsion along
with a shift of $60^{\circ}$. Statistical errors on the values obtained
from MD as well as on the calculated RMSE are less than $0.1Hz$.\label{tab:-adenosine-scalar-coupling}}
\end{table}
 As it can be seen, the RMSE is greatly reduced compared to the original
Amber force field. Lagrangian multipliers are shown in Supporting
Information  (table S2). We recall that the greater the value of
$\sigma$ the higher the confidence in the force field and the lower
the correcting MaxEnt potential. A plot of RMSE vs $\sigma$ is provided
in Supporting Information (figure S3). We notice that in this case
an arbitrary small RMSE can be obtained by choosing a negligible value
of $\sigma$. It should be noticed that enforcing a RMSE smaller than
the typical RMSE between different set of parameters in Karplus relations ($\approx 0.6$ Hz)
is not meaningful. Moreover, this would introduce much larger corrections
to the force field  (see Supporting Information figure S4) that could
lead to uncontrolled artifacts. For instance, in some of the simulations
using $\sigma=0$ we obtained stereoisomerizations of the C2' atom
of the sugar  (data not shown). With the adopted value of $\sigma\ =\ 2$
the effect of the corrections on the one-dimensional free-energy profiles
of the refined dihedral angles is $\leq2\ K_{b}T$. Free-energy profiles
for a set of representative torsional angles are shown in \figref{adenosine-free-energy}.

\subsection{Using $^{3}J$-coupling for self-consistent force-field refinement}

We use the introduced procedure to perform a self-consistent force-field
refinement on a set of RNA nucleosides and dinucleoside monophosphates.
In \figref{torsionals}b the ApC dinucleoside is shown 
and all the torsions considered in the refinement procedure are indicated. The obtained
Lagrangian multipliers for each torsional angle are summarized in
\tabref{Lagrangian-multipliers-karplus-selfcons}. When fitting systems
involving different nucleobases  (e.g A and C), torsions around the
glycosidic bond were considered as base dependent, together with the
$\nu_{3}$ torsion, which we empirically observed to be the sugar
torsion that is most correlated with the base/sugar relative orientation.
Such torsions will feel a different correcting potential depending
on whether they belong to and Adenosine or Cytosine. Base dependent
torsions are highlighted in red in \figref{torsionals}b. In case
of a duplicated term in a single simulation  (e.g., the $\chi$ angle
in an adenine which appears twice in the ApA dinucleoside monophosphate),
we do not enforce their individual values but the sum of the two scalar
couplings to match the sum of the corresponding experimental values.

\begin{figure}
\begin{centering}
\includegraphics[scale=1.5]{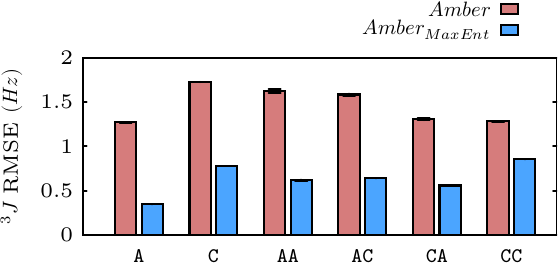}
\par\end{centering}

\caption{$^{3}J$ RMSE for each system with the Amber force-field and the Amber$_{MaxEnt}$
force-field obtained with the self consistent refinement.\label{fig:-RMSE-self_cons}}

\end{figure}

 RMSEs for each system are shown in \figref{-RMSE-self_cons}. Here
it can be appreciated that all the resulting RMSEs are below 1 Hz.
We notice that in this case the number of non-equivalent dihedrals
 (16) is significantly lower than the number of experimental data
 (78). This means that data are redundant and the procedure can detect
potential inconsistencies between experimental data.

Back calculated $^{3}J$ couplings for each torsion and Karplus parameters
are provided in Supporting Information  (table S3,S4,S5). The effect
of the corrections on the one-dimensional free-energy profiles associated
with all the dihedral angles is shown in Supporting Information (figure
S10,S11)

\begin{table}
\begin{centering}
\begin{tabular}{|ccc>{\centering}m{2.1cm}|}
\hline 
\multicolumn{1}{|c|}{Coupling} & \multicolumn{1}{c|}{Torsion $\theta$} & \multicolumn{1}{c|}{Base} & Lagrangian

multiplier ($Hz^{-1}$)\tabularnewline
\hline 
$^{3}J_{H1^{\prime}H2^{\prime}}$ & {\scriptsize{}$\nu_{1}$} & \texttt{A},\texttt{C} & 0.4393\tabularnewline
\hline 
$^{3}J_{H2^{\prime}H3^{\prime}}$ & {\scriptsize{}$\nu_{2}$} & \texttt{A},\texttt{C} & 0.0570\tabularnewline
\hline 
\multirow{2}{*}{$^{3}J_{H3^{\prime}H4^{\prime}}$} & \multirow{2}{*}{{\scriptsize{}$\nu_{3}$}} & \texttt{A} & 0.4009\tabularnewline
 &  & \texttt{C} & 0.3316\tabularnewline
\hline 
$^{3}J_{H4^{\prime}H5^{\prime}}$ & {\scriptsize{}$\gamma'$} & \texttt{A},\texttt{C} & 0.3643\tabularnewline
\hline 
$^{3}J_{H4^{\prime}H5^{\prime\prime}}$ & {\scriptsize{}$\gamma''$} & \texttt{A},\texttt{C} & -0.2077\tabularnewline
\hline 
$^{3}J_{H3^{\prime}P}$ & {\scriptsize{}$\epsilon_{1}$} & \texttt{A},\texttt{C} & -0.2358\tabularnewline
\hline 
$^{3}J_{H5^{\prime}P}$ & {\scriptsize{}$\beta_{2}$} & \texttt{A},\texttt{C} & -0.0237\tabularnewline
\hline 
$^{3}J_{H5^{\prime\prime}P}$ & {\scriptsize{}$\beta_{2}$} & \texttt{A},\texttt{C} & -0.0700\tabularnewline
\hline 
$^{3}J_{C2^{\prime}P}$ & {\scriptsize{}$\epsilon_{1}$} & \texttt{A},\texttt{C} & 0.2015\tabularnewline
\hline 
\multirow{2}{*}{$^{3}J_{C4^{\prime}P}$} & {\scriptsize{}$\epsilon_{1}$} & \texttt{A},\texttt{C} & 0.2010\tabularnewline
 & {\scriptsize{}$\beta_{2}$} & \texttt{A},\texttt{C} & 0.1923\tabularnewline
\hline 
$^{3}J_{H1^{\prime}C4}$ & \multirow{2}{*}{{\scriptsize{}$\chi$}} & \texttt{A} & 0.1758\tabularnewline
$^{3}J_{H1^{\prime}C2}$ &  & \texttt{C} & 0.4270\tabularnewline
\hline 
$^{3}J_{H1^{\prime}C8}$ & \multirow{2}{*}{{\scriptsize{}$\chi'$}} & \texttt{A} & -0.4068\tabularnewline
$^{3}J_{H1^{\prime}C6}$ &  & \texttt{C} & -0.7401\tabularnewline
\hline 
\end{tabular}
\par\end{centering}

\caption[Lagrangian multipliers in the self consistent procedure and Karplus
relations]{Lagrangian multipliers associated to each torsional angle used in
the self consistent procedure together with the associated Karplus
parameters used to back calculate $^{3}J$ scalar couplings. The third
column specifies to which system the corrections have to be applied.
Karplus relations used are in the form $^{3}J(\theta)=A\cos^{2}(\theta+\varphi)+B\cos(\theta+\varphi)+C\sin(\theta+\varphi)\cos(\theta+\varphi)+D$.
$\chi'$ is defined as $H1^{'}-C1^{'}-N1/N9-C6/C8$ along with a phase
shift of $60^{\circ}$. \label{tab:Lagrangian-multipliers-karplus-selfcons} }
\end{table}

\subsection{Validation on RNA Tetranucleotides}

\begin{figure}
\begin{centering}
\includegraphics[scale=2]{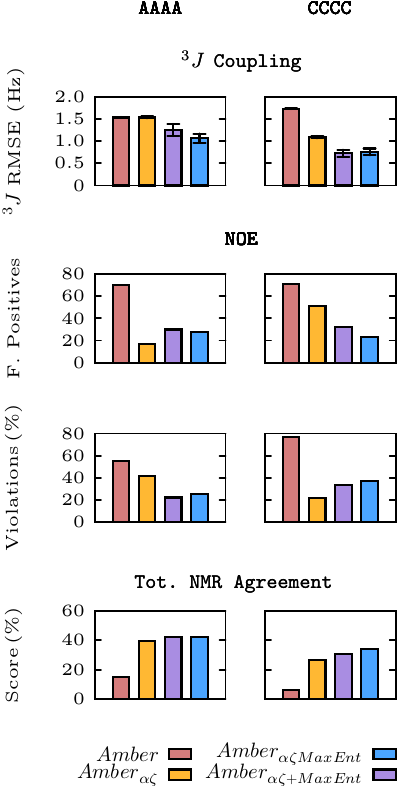}
\par\end{centering}

\caption{Agreement with the NMR solution experiments for Amber, Amber$_{\alpha\zeta}$
and Amber$_{\alpha\zeta MaxEnt}$. The number of distance false positives represent the MD predicted NOEs not observed in the experiments.\label{fig:tetranucl_valid}}

\end{figure}

\begin{figure}
  \begin{centering}
    \includegraphics[scale=1.5]{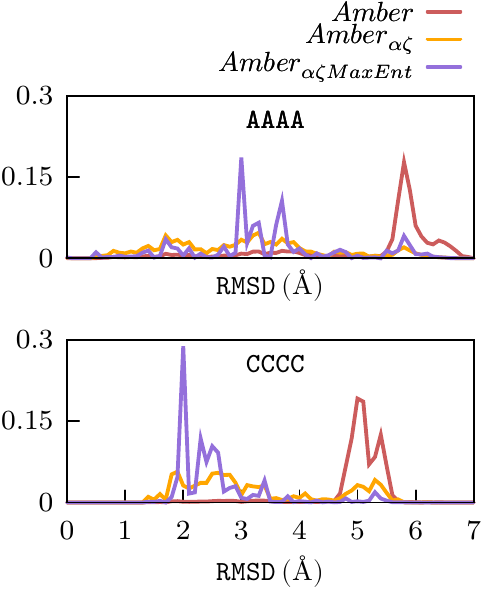}
  \par\end{centering}

  \caption{Structural ensembles obtained with Amber, Amber$_\alpha \zeta$ and Amber$_{\alpha\zeta MaxEnt}$.\label{fig:tetranucl_ensembles}
Ensembles are represented by showing the histogram of the RMSD from the ideal A-form conformation. Both Amber$_\alpha \zeta$ and
Amber$_{\alpha\zeta MaxEnt}$ show a significant decrease in the population of the high RMSD structures which are associated to intercalated 
conformations.}

\end{figure}

The corrected force field is then validated on two RNA tetranucleotides,
AAAA and CCCC. In a previous work\cite{Gil-LeyBottaroBussi2016} we
have shown that on such systems a significant improvement of the agreement
with NMR solution experiments can be obtained penalizing $\alpha(g+)/\zeta(g+)$
conformations. These conformations are associated to intercalated
structures\cite{CondonTurner2015,Gil-LeyBottaroBussi2016} that are
incompatible with solution experiments. We call here Amber$_{\alpha\zeta}$
a potential obtained adding to Amber a two dimensional Gaussian potential
centered on the $\alpha(g+)/\zeta(g+)$ conformation with a standard
deviation of $0.7$ rad and height $8\ \frac{kJ}{mol}$. The Lagrangian
multipliers discussed above were obtained as corrections to be applied
on the Amber force field. We here perform a new self-consistent fit
with identical simulation parameters using as prior distribution the
Amber$_{\alpha\zeta}$ potential and call Amber$_{\alpha\zeta MaxEnt}$
the resulting force field. We also define the Amber$_{\alpha\zeta+MaxEnt}$
force field as the one obtained by adding the corrections obtained
in the previous section on top of the Amber$_{\alpha\zeta}$ force-field,
without repeating the self-consistent refinement. In order to asses
the performance of Amber, Amber$_{\alpha\zeta}$, Amber$_{\alpha\zeta+MaxEnt}$
and Amber$_{\alpha\zeta MaxEnt}$ we performed the same analysis as
in refs\cite{CondonTurner2015,Gil-LeyBottaroBussi2016} on AAAA and
CCCC. This analysis is made by reweighting the trajectories described
in ref \cite{Gil-LeyBottaroBussi2016}.  
For each force field, we evaluate
the RMSE associated to scalar coupling as well as the number of violations
and false positives in contacts predicted by nuclear Overhauser experiments
 (NOEs). NOEs are particularly important in tetranucleotides since
they are sensitive to intercalated structures erroneously obtained
using the Amber force field that have been previously reported \cite{CondonTurner2015,BergonzoCheatham2015,BottaroGil-LeyBussi2016,Gil-LeyBottaroBussi2016}.
We notice that NOEs might not be visible for many reasons others than the distance is too large.
This often happens with large RNAs and proteins and can be due to
(1) one or both of the involved resonances are broader than others due to local conformational flexibility at an intermediate rate (microsecond to millisecond), or (2) chemical exchange with solvent protons. All the observed signals in these small systems have similar linewidths (i.e. no intermediate conformational exchange) and only non-exchangeable protons are analyzed.
Additionally, for a similar tetranucleotide (GACC) it was shown that intercalated structures would
lead to peaks that would be easy to detect because they would appear in unique and uncrowded regions
of the spectra.\cite{Yildirim2011}
Comparison of MD with NMR for the tetranucleotides is reported in \figref{tetranucl_valid}.
As it can be seen,
the MaxEnt corrections improve the agreement with experimental scalar
couplings for AAAA and CCCC with respect to both Amber and Amber$_{\alpha\zeta}$
force fields. When considering the NOEs, it can be appreciated that
the largest improvement with respect to Amber originates from the
$\alpha\zeta$ correction, as previously suggested. Interestingly,
the MaxEnt corrections further decrease the number of false positives
in CCCC and the number of violations in AAAA. We summarize the agreement
with experimental NOEs using the NMR score defined in Ref \cite{CondonTurner2015}.
When comparing Amber$_{\alpha\zeta MaxEnt}$ with Amber$_{\alpha\zeta+MaxEnt}$
it can be noticed that performing a new self-consistent fit starting
from Amber$_{\alpha\zeta}$ represent a better choice since it improves
both the RMSE and the total NMR agreement.  We remark that this is
a completely independent validation since experimental data for AAAA
and CCCC were not considered in the self-consistent force-field refinement
procedure. Moreover, we stress that the validation is made on systems
that are different from those used in the fitting procedure. This
suggests the corrections to be portable to larger RNA molecules. 
We finally notice that if the magnitude of the correcting potential is larger than a few $k_BT$
the reweighting procedure can lead to very poor sampling\cite{shen2008statistical,ceriotti2011inefficiency}. To assess the confidence in the reweighting we computed both the Kish's effective sample size\cite{GrayKish1969} and the statistical error on the RMSE. The Kish's effective sample sizes are respectively 10 (CCCC) and 29 (AAAA)
for the Amber$_{\alpha\zeta MaxEnt}$ potential, to be compared to 4000 frames in the unbiased trajectories. Despite these numbers might seem low,
the impact of the reweighting procedure on the estimated RMSE is better described by its statistical error.
Although the statistical error is significantly increased in the reweighted ensemble (see \figref{tetranucl_valid}), its value is still small enough to allow
for a proper comparison between the RMSEs.
The structural ensembles obtained with Amber, Amber$_\alpha \zeta$ and Amber$_{\alpha\zeta MaxEnt}$ are also shown in \figref{tetranucl_ensembles}. It can be appreciated that in both AAAA and CCCC the effect of the MaxEnt corrections is to penalize structures with high value of root-mean-square deviation (RMSD)
after optimal superposition from the ideal A-form conformation, which are related to wrongly predicted intercalated conformations.

\section{Discussions}

In this paper we introduced a framework to enforce on the fly noisy
data from bulk experiments on molecular dynamics simulations. In the
first part (see subsection 1 of Method) we discussed the case of experiments
without noisy tolerance. This procedure is completely equivalent to
the MaxEnt procedure discussed by Chodera and Pitera \cite{PiteraChodera2012}
and share many similarities with the experimentally directed simulation
(EDS) introduced by White and Voth\cite{WhiteVoth2014}. In particular,
the only difference between the implementation of the MaxEnt procedure
used here and EDS is that we here used a different optimization procedure
to find the Lagrangian multipliers. In the second subsection of Method
we extend the previous approach so as to take into account experimental
uncertainties. Several Bayesian approaches have been discussed to
model experimental errors in similar contexts  (see e.g. \cite{Rieping2005,OlssonFrellsenBoomsmaEtAl2013,Hummer2015,BrookesHead-Gordon2016,BonomiMETAINF2016}).
Methods have been described to reweight a pre-computed ensemble of
structures so as to match experimental averages \cite{Das2014,Hummer2015,BrookesHead-Gordon2016}.
We here apply the MaxEnt procedure on an extended system where fictitious
variables are introduced that take into account the discrepancy between
theory and experiment. A suitably chosen prior distribution for these
variables allows one to control the accuracy of the fitting and to
embed in the calculation the confidence in the original force field. 

The procedure is iterative and is completely encoded in the update
rule stated in Equation \ref{eq:update-rule}. It is important to
notice that a similar equation could be obtained using theoretical
approaches different from the one introduced in this paper. For instance,
one could decide to maximize the posterior as a function of the residuals
$\epsilon$ as it is done in Ref\cite{Hummer2015}, instead of computing
their average value. More comments on this analogy can be found in
Supporting Information (section 3).

We notice that other methods have been proposed in the past to model
noisy data within the MaxEnt framework. For instance, Chen and Rosenfeld\cite{Chen99agaussian}
have proposed to introduce a Gaussian prior on the Lagrangian multipliers.
The Laplace prior on the additional variables used here has a similar
effect, and allows the range of values for the Lagrangian multipliers
to be explicitly controlled.

An alternative formulation of the MaxEnt procedure discussed here
can be obtained by replacing the time averages with averages performed
on an ensemble of molecular dynamics simulations\cite{BestVendruscolo2004,LindorffLarsenBestVendruscolo2005}.
Replica averaging only converges to MaxEnt when an infinite number
of replicas is simulated\cite{CavalliCamilloniVendruscolo2013,RouxWeare2013}
and implies an intrinsic statistical error in the averages when used
with a finite number of replicas. Replica formalism has been extended
so as to take into account experimental errors\cite{Hummer2015,BonomiMETAINF2016}.
In this context, we preferred to use an iterative procedure since
it allows Lagrangian multipliers to be estimated on the fly. The statistical
error that in our procedure arises from the finite length of the simulation
can be assessed by standard blocking analysis.

The tests that we performed on a model one-dimensional system with
a bi-modal distribution allow to easily understand the effects of
the chosen parameters on the resulting ensembles. In particular, the
variance of the prior distribution used for the additional variables
can be used to tune the relative weight of the original model and
of the enforced experimental data. A Laplace prior for these variables
allows for outliers to be tolerated. 

We then applied the method to an important open problem, that is the
refinement of a force field in order to reproduce available NMR data
for RNA oligomers. At first we use our method to enforce all the $^{3}J$
scalar couplings available for the four nucleosides. Since the free-energy
landscape of nucleosides have significant barriers, we here combined
the approach with an enhanced-sampling method based on multiple replicas.
This can be straightforwardly done in our formulation since the on-the-fly
estimated Lagrangian multipliers can be instantaneously transferred
to the biased replicas. The results display a significantly reduced
RMSE with respect to experimental data when compared to the original
Amber force field. This is expected, since the validation is made
against the same dataset used for the training. However, this confirms
that the methodology converges to the correct result also in a non
trivial model system.
We also observe that the employed couplings are unevenly distributed
along the RNA backbone. If desired, one could associate a lower value
of $\sigma$ to the individual couplings that are considered more relevant so as to
increase their weight in the fitting procedure.

The method is then applied to the self-consistent force-field fitting
for two RNA nucleotides  (A and C), employing a variety of data measured
for several systems  (A and C nucleosides, as well as ApA, ApC, CpA,
and CpC dinucleoside monophosphates). Also here, the procedure takes
implicitly advantage of the on-the-fly transferability of the Lagrangian
multipliers. Our approach reminds the spirit behind the restrained
ESP charge model\cite{resp1993}, where equivalent atoms are restrained
to have equivalent charges. This is translated here in having same
correcting potentials on chemically equivalents dihedrals independently
of their position in the sequence. Notice that using a self-consistent
procedure where several terms are restrained to be identical effectively
reduces the flexibility of the resulting force field and implicitly
decreases its capability to match the experimental data. For instance,
in the case of a duplicated term in a single simulation  (e.g., the
$\chi$ angle in an adenine which appears twice in the ApA dinucleoside
monophosphate), our approach is only controlling the sum of the two
scalar couplings and not their individual values. In our specific
application, the number of independent parameters in the force field
is 16, which should be compared with 78 independent experimental data.
In this respect, it is important to notice that in this application
the calculation of the RMSE, which depends also on the non-explicitly
controlled observables, allows for a rigorous cross validation of
the method.

The functional form of the corrections derived here, which is proportional
to the Karplus equations, is compatible with the one of dihedral potentials.
This suggests the use of scalar coupling data as an alternative to
quantum chemistry calculations for force-field parametrization
or as a refinement tool on top of quantum-chemistry derived torsions.
One might
be concerned about the fact that corrections developed to match experimental
data on small systems are not necessarily portable to larger systems.
However, it must be observed that the standard procedure used in the
Amber force field is to refine dihedral potentials based on quantum
chemistry calculations performed on small fragments, whose typical
size is often below the size of the systems considered in this work\cite{ZgarbovaOtyepkaSvponerEtAl2011,IvaniDansNoyEtAl2015}.
As a validation, we performed a reweighting of previously published
trajectories for two RNA tetranucleotides  (AAAA and CCCC). In spite
of their apparent simplicity these unstructured oligomers are not
described properly by any of the current versions of the Amber force
field \cite{BergonzoCheatham2015}. Our results show that the corrections
are portable and significantly improve the description of these tetranucleotides.
The resulting RMSEs are below 1 Hz, which is the typical difference
between alternate Karplus equations. The development of a force field
that consistently describes all nucleotides and dinucleosides, as
well as its validation on tetranucleotides and larger systems, is
left as a subject for a future investigation.

In conclusion, we introduced a novel procedure that allows experimental
errors to be explicitly modeled in a MaxEnt framework. The method
is applied to the self-consistent force-field fitting on RNA systems.
Results indicate that the obtained force-field corrections are portable
and suggest a new paradigm for empirical force-field refinement.

\acknowledgement

Massimiliano Bonomi, Carlo Camilloni, Michele Parrinello, Omar Valsson,
Gregory Voth, and Andrew White are acknowledged for carefully reading
the manuscript and providing several useful suggestions.
Doug Turner and Scott Kennedy are acknowledged for useful discussions on the interpretation
on NOE spectra.
The research
leading to these results has received funding from the European Research
Council under the European Union\textquoteright{}s Seventh Framework
Programme (FP/2007-2013) / ERC Grant Agreement n. 306662, S-RNA-S.

\begin{suppinfo}
Three sections discussing (1.1) the case of correlated observables, (1.2) the MaxEnt algorithm scheme, (1.3) a family of priors for modelling errors, (1.4) an extension of the method to the case of many restraints with the same error, (1.5) the self-consistent fit algorithm scheme, (1.6) the RMSD between different Karplus parameters, (2) a one dimensional model system to highlight the features of different error models and (3) a comparison of MaxEnt and maximum a posteriori. Plots of different prior functions (S.1); one dimensional model system (S.2); RMSE as function of $\sigma$ (S.3); correcting potential on Adenosine $\chi'$ torsion (S.4); sample Plumed input file for Adenosine (S.5); sample Plumed input files for the self-consistent procedure for Adensosine, ApA and Apc (S.6, S.7, S.8, S.9); free-energy profiles in the refined $Amber_{\alpha \zeta MaxEnt}$ force field for A, C, ApA, ApC, CpA, CpC (S.10, S.11); comparison of maximum entropy and Maximum a posteriori (S.12); $^3$J scalar couplings and Lagrangian multipliers for A, G, C, U (Tables S.1, S.2); $^3$J scalar couplings and Lagrangian multipliers for A, C, ApA, ApC, CpA, CpC  in the self consistent procedure (Tables S.3, S.4, S.5); standard deviation between $^3$J scalar coupling using various sets of Karplus parameters.
\end{suppinfo}

\providecommand{\latin}[1]{#1}
\providecommand*\mcitethebibliography{\thebibliography}
\csname @ifundefined\endcsname{endmcitethebibliography}
  {\let\endmcitethebibliography\endthebibliography}{}

\begin{tocentry}
  \includegraphics{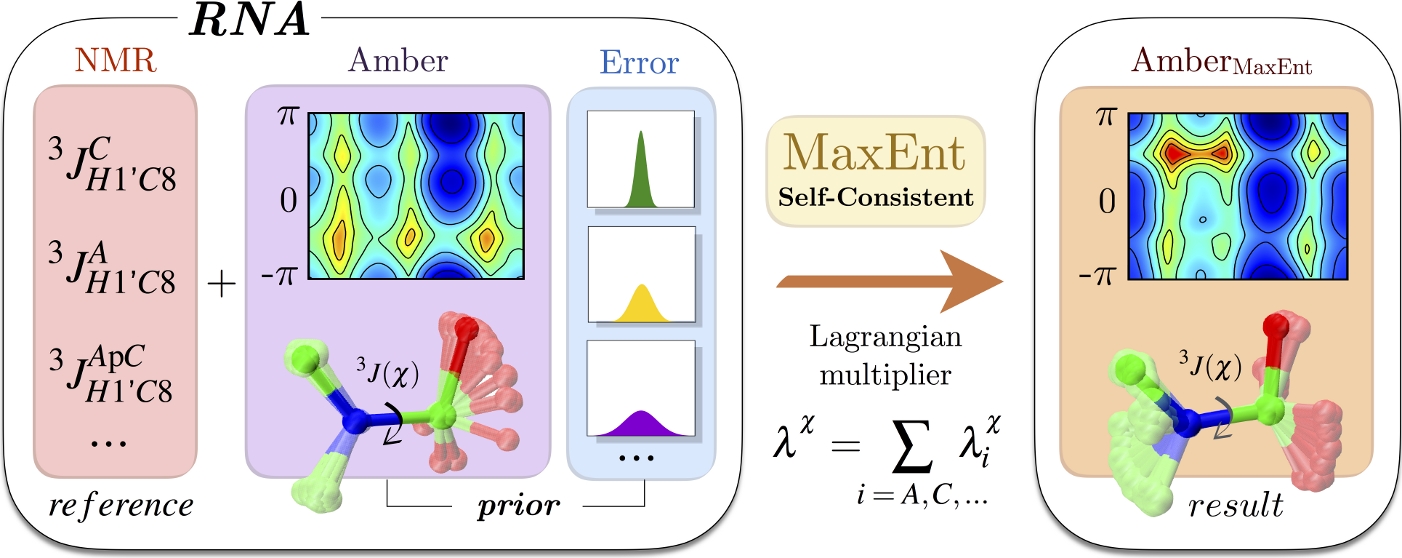}
  \label{For Table of Contents Only}
\end{tocentry}
\end{document}